\documentclass[fleqn,twoside]{article}
\usepackage{espcrc1}
\usepackage{graphicx}
\usepackage[figuresright]{rotating}

\newcommand{\AmS}{{\protect\the\textfont2
  A\kern-.1667em\lower.5ex\hbox{M}\kern-.125emS}}
\title{\bf Study of CR primaries and their cascades at $E_\circ$ = 10 $\div$ 100 TeV through EAS-TOP and MACRO}
\author{The EAS-TOP and MACRO Collaborations
\thanks{The complete list of authors will be reported in the summary of the 
TAUP2001 Proceedings volume.}`}

\begin{document}
\maketitle

\begin{abstract}
A measurement of the lateral distribution of Cherenkov light in Extensive Air
Showers (EAS) at $E_\circ$ = 10 $\div$ 100 TeV and a study of 
the compatibility of the photon number spectra with the expectations from
the direct measurements of $p$ and $\alpha$ spectra and the Corsika-QGSJET
propagation code in the atmosphere
have been performed at the National Gran Sasso
Laboratories by the EAS-TOP and MACRO arrays. The telescope array of 
EAS-TOP 
has been used as the Cherenkov light detector. The muon tracking system of
MACRO in the deep underground Gran Sasso Laboratories ($E_{\mu}$ $>$ 1.3 TeV)
served as the EAS detector, including core localization and arrival direction. 
\end{abstract}

\section{INTRODUCTION}

The EAS-TOP
and MACRO arrays at the Gran Sasso Laboratories offer a unique opportunity of 
measuring
the lateral distribution of Cherenkov light in the 10 $\div$ 100 TeV energy
range by associating the Cherenkov light collected by the EAS-TOP telescopes 
with 
the TeV muon reconstruction, and consequently the EAS core geometry, through
the MACRO array. 
In this paper we report on the  
measurements of the Cherenkov light lateral distribution compared 
with the 
results of simulations based on the 
CORSIKA-QGSJET code providing an experimental validation of the code itself.
Moreover the technique
allows a study of the primary composition and a comparison with the direct 
existing measurements in a overlapping region. 
Due to the shower selection through the high energy
muon ($E_{\mu} >$ 1.3 TeV, i.e. primary energy $E_{o} >$ 1.3 TeV/nucleon), 
in the energy
range 10 TeV $<$ $E_{o}$ $<$ 40 TeV (10 TeV being the Cherenkov 
telescopes' threshold energy) the selected primaries are mainly protons, while
for 40 TeV $<$ $E_{o}$ $<$ 100 TeV they include both $p$ and $\alpha$
particles.\\

\section{DETECTORS AND DATA REDUCTION}

The Cherenkov array of EAS-TOP (\cite{Agl1eBer2}) consists of 7 telescopes
60-80 m apart from each other. Each telescope loads two wide angle detectors
equipped with 7 photomultipliers (PMs) (d $=$ 6.8 cm each) on the focal plane 
of
parabolic mirrors (0.5 m$^2$ area, 40 cm focal length) for a total field of
view (f.o.v.) of 0.16 sr.\\
MACRO, in the underground Gran Sasso Laboratories at 963 m a.s.l., 3100 m w.e.
of minimum rock overburden, is a large area multi-purpose apparatus designed
to detect penetrating cosmic radiation. 
A detailed description of
the apparatus can be found in \cite{MAC3e4}. In this work we consider muon
tracks, having at least 4 aligned hits in both views of
the horizontal streamer tube planes over the 10 layers composing the whole
detector.\\ 
The two experiments are separated by a thickness of rock ranging from 1100 m
up to 1300 m, depending on the angle.

The corresponding minimum energy for a muon to reach the depth of MACRO ranges
from 1.3 to 1.8 TeV. Event coincidence is established off-line,
using the absolute time provided by a GPS system with an accuracy better than
1 $\mu$s.\\
The two experiments have run in coincidence in the bright moonless nigths in 
the period 1998 - 2000. Here we report on the analysis 
performed using 5 telescopes
in coincidence for a live time $\Delta$T = 208 hours corresponding to an
exposure $\Gamma \approx 815$ day $\cdot$ m$^2$ $\cdot$ sr. In such period 
MACRO 
reconstructed
35814 events in the angular field 16$^{\circ}$ $<$ $\theta$ $<$ 58$^{\circ}$ 
and 127$^{\circ}$ $<$
$\phi$ $<$ 210$^{\circ}$, corresponding to the region in zenith and azimuth
covered by the Cherenkov telescopes. 3830 events have been found in coincidence
with Cherenkov data in a window of $\Delta t =$ 7$\mu$s, the expected 
accidental contamination being 3.0 events.\\
From the point of view of the muon reconstruction, the standard MACRO procedure
\cite{MAC3e4} provides an accuracy of 0.95$^{\circ}$ (due to instrumental
uncertainties and the scattering in the rock) that combined with the muon 
lateral spread leads to an uncertainty on the EAS core location of 
$\Delta x_c$ $\approx$
20 m.\\
Concerning Cherenkov light, the data treatment is summarized in 
\cite{Agl1eBer2} 
where sky luminosity and mirror reflectivity's variations, PMs' gain 
calibration, absolute normalization among PMs, 
photoelectron - photon conversion and light collection efficiency are taken into
account. Considering all the different components, a systematic error of
$\sigma_{sys} \approx$ 21\% has been evaluated.  

\section{THE SIMULATION}

The Cherenkov light lateral distribution was calculated from simulated
showers generated with the CORSIKA code version 5.61 (\cite{Kna5})
and QGSJET hadron interaction model. Both protons and Helium nuclei were
considered as primary particles, with discrete energies between 20 and 120 TeV.
Zenith and azimuth angles were chosen randomly inside the telescopes' fields of
view (30$^{\circ}$ $<$ $\theta$ $<$ 40$^{\circ}$ and 175$^{\circ}$ $<$ $\phi$
$<$ 185$^{\circ}$). The requirement of having a muon with energy $E_{\mu}$ $>$
1.3 TeV implies a reduction of $\approx$ 20\% of the absolute photon densities
and this effect has been taken into account. Proton and Helium lateral 
distributions show a similar shape with a 10 - 20 \% difference in intensity
depending on the energy.  

\section{THE ANALYSIS}

The lateral distribution was constructed using the constant
intensity cut technique (c.i.c.) \cite{Agl1eBer2}. 
Photon integral spectra corresponding to 6 different coronae (r $\in$ [0,20], 
[20,35],
[35,50], [125,145], [145,165], [165,185] m) where the
central or the lateral PMs were fully efficient have been considered for each
telescope, normalized in area and time and summed up with the corresponding 
ones of all telescopes. 

\begin{figure}[htb]
\begin{center}
\vspace{-1.cm}
\includegraphics[width=12cm]{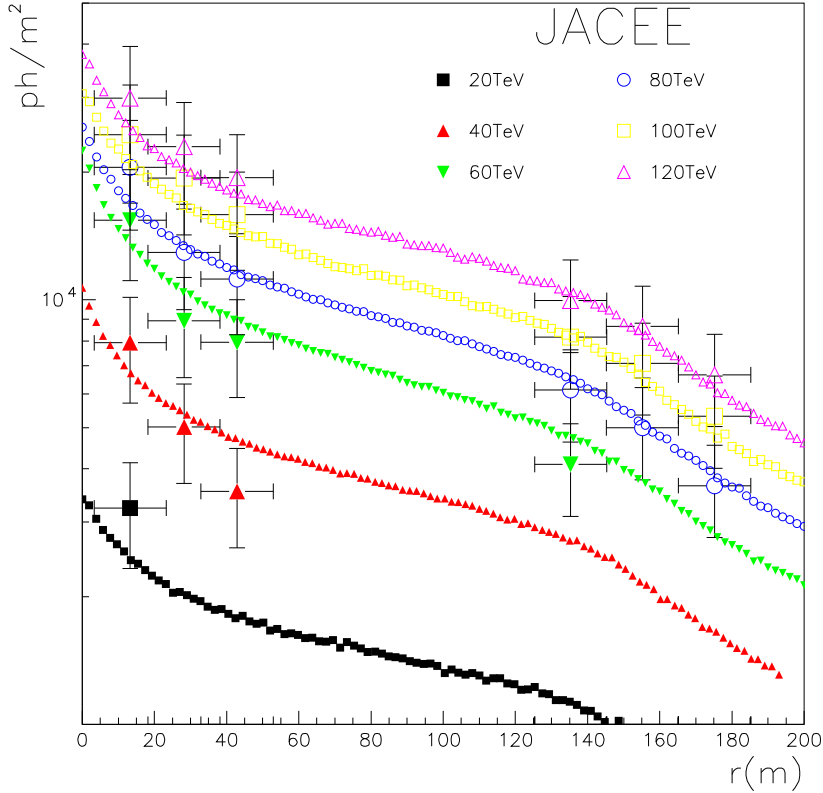}
\vspace{-1.0cm}
\caption{Measured C.l. lateral distributions compared with simulated ones
(290 $<$ $\lambda$ $<$ $630$ nm) using the JACEE spectra.
\label{fig:jacee}}
\end{center}
\end{figure}

\begin{figure}[htb]
\begin{center}
\vspace{-1.cm}
\includegraphics[width=12cm]{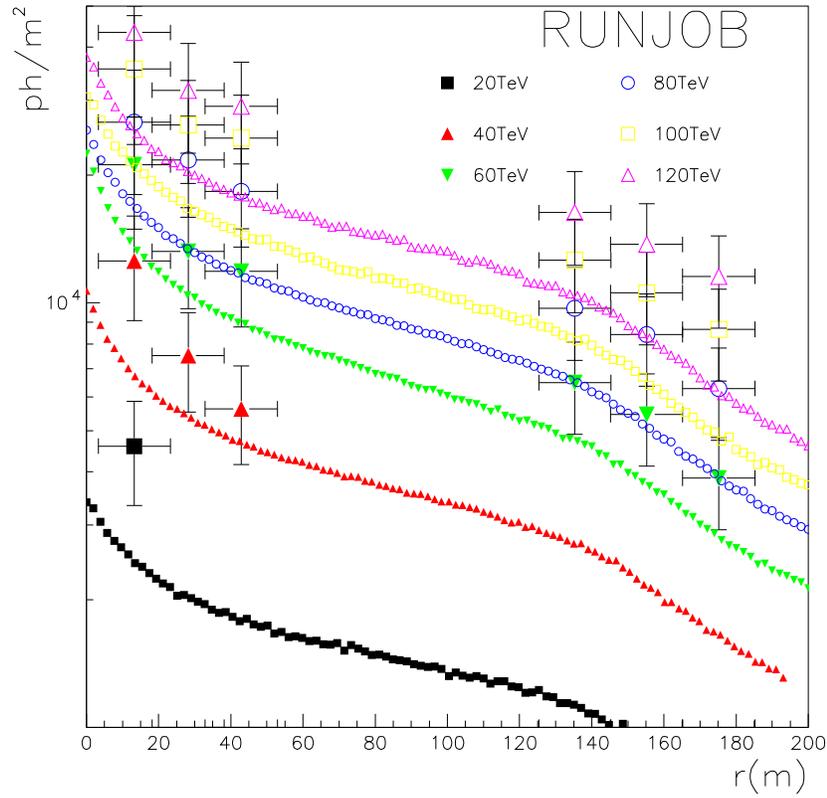}
\vspace{-1.0cm}
\caption{Same as fig.~\ref{fig:jacee} using the RUNJOB spectra.
\label{fig:runjob}}
\end{center}
\end{figure}

The number of photons corresponding to the same rate in the
6 different coronae was used to construct a lateral distribution.\\
Frequencies were selected according to the following expression:
\begin{eqnarray}
f \left ( E > E_o \right ) = \int_{E_o}^{\infty} \frac{dN_p}{dE} \cdot
p_{p}^{\mu}(E)dE + \nonumber \\  
\int_{E'_o}^{\infty} \frac{dN_{He}}{dE'} \cdot p_{He}^{\mu}(E')dE'
\end{eqnarray}
where E$_o$ and E'$_o$ represent respectively the proton and
helium energies giving the same lateral distribution. The following values
have been chosen for protons: 20, 40, 60, 80, 100 and 120 TeV.
$\frac{dN_p}{dE}$ and $\frac{dN_{He}}{dE}$ are the differential primary
spectra: the JACEE and RUNJOB data have been used as reported in 
tab.~\ref{table:spe}. 
\vspace{-0.5cm}
\begin{table}[!htb]
\caption{JACEE and RUNJOB \cite{JAC6eRUN7}
primary spectra.}
\label{table:spe}

\begin{tabular}{lll}
\hline
el.&JACEE&RUNJOB\\
\hline
$p$&$0.111 \times E^{-2.8}$&$0.126 \times E^{-2.8}$\\
$He$&$7.86 \cdot 10^{-3} \times E^{-2.68}$&$4.42 \cdot 10^{-3} \times
E^{-2.8}$\\
\hline
\end{tabular}\\[2pt]
Units are:
m$^{-2}$s$^{-1}$sr$^{-1}$TeV$^{-1}$/n$^{-1}$.
\end{table}
\vspace{-0.5cm}
Finally $p_{p}^{\mu}$ and $p_{He}^{\mu}$ represent the probability
for a primary $p$ or $\alpha$ to produce a muon with energy $E_{\mu}$ $>$ 
1.3 TeV in the MACRO detector. Such contribution has been calculated
through 
the CORSIKA-QGSJET code in the atmosphere and  
detector using the Gran Sasso Interface program (CORGSI) for muon propagation
in the rock. 
As it can be seen from Fig.~\ref{fig:jacee}, the experimental points match very
 well with the simulated ones according to the JACEE proton and helium spectra.
The error on the $x$ axis represents the uncertainty on the EAS core
position, while on the $y$ axis statistical and systematic errors are summed up
in quadrature. The systematic error is of the order of 20\% and its effect is
to scale all the curves without changing their shape.\\
The C.l. yield associated with the underground muon 
reconstruction provides a
technique to discriminate among different primary spectra. 
In fact the agreement is worse when
frequencies are calculated using RUNJOB spectra of Tab.~\ref{table:spe} 
(see 
Fig.~\ref{fig:runjob}). This has to be
ascribed to the lower contribution of the $\alpha$ component in the RUNJOB 
spectra.

\section{CONCLUSIONS}

A measurement of the lateral distribution of Cherenkov light in EAS in the 
energy range 20 $\div$ 120 TeV has been performed at the
Gran Sasso Laboratories by the EAS-TOP and MACRO arrays. The EAS and its 
geometry are selected through the muon detected deep underground by MACRO
(E$_{\mu}$ $>$ 1.3 TeV). The measurements are performed by means of the 
Cherenkov light detector of EAS-TOP at Campo Imperatore (2000 m a.s.l.).
The measurement is compared with the results of simulations based on the
CORSIKA-QGSJET code. Simulated and real data show a good agreement, inside
20\% systematic uncertainties.\\
The shape of the l.d.f. reflects the rate of energy release in the atmosphere,
(i.e. the properties of the interaction, the primaries being 
dominated by the lightest components due to the TeV muon trigger requirement)
while the absolute scale is mostly related to the event rate, i.e. the primary
$p$ and $\alpha$ spectra. The agreement of both of them 
(see fig.~\ref{fig:jacee}) shows both the adequacy of the CORSIKA-QGSJET code
in describing the cascades in this energy range and of the JACEE flux in the 
20 $\div$ 120 TeV region.
The contribution of fluctuations and of the CNO component have been 
successively studied and
they do not affect these conclusions.

\end{document}